\documentclass[12pt,nofootinbib,nopacs, amssymb]{revtex4}
\begin{document}
\title{Angular Functions with Complex Angular Momenta
\footnote{Published in Soviet Journal of Nuclear Physics,
Volume 4, Number 3, March 1967, pp.469-475.
Translated by W.H. Furry from the Russian original
Yad. Fiz. \textbf{4}, 663-672 (September, 1966).
Present address (2013): Petersburg Nuclear Physics Institute;
e-mail: azimov@thd.pnpi.spb.ru. The published English text is
reproduced here mainly as it was, with numerous misprints
corrected; all the footnotes have been appended in 2013.}}

\author{Ya.~I.~Azimov$^{\,a}$}

\affiliation{$^a$A. F. Ioffe Physico-Technical Institute,
Academy of Sciences of the U.S.S.R. \\
Submitted to JNP editor February 19, 1966\\
J. Nucl. Phys. (U.S.S.R.) \textbf{4}, 663-672
(September, 1966) }

\bigskip

\begin{abstract}

In the study of the amplitudes for many-particle processes,
and also for processes involving particles with spin, the
use is made of matrix elements of the rotation group
$d^j_{\mu\nu}(z)$. In this paper the generalization of the
functions $d^j_{\mu\nu}(z)$ to arbitrary arguments and
indices is studied. At the same time the functions of the
second kind, analogous to Legendre functions of the second
kind, are investigated. The results obtained play an important
part in the introduction of complex angular momenta in
many-particle processes. [Added in 2013: The generalized
Legendre functions considered here may be applied as well to
many other problems.]

\end{abstract}


\maketitle


The study of the partial amplitudes for elastic scattering of two
spinless particles for complex orbital angular momenta requires a
knowledge of the properties of the Legendre functions $P_j(z)$
and $Q_j(z)$ for arbitrary values of the argument and index. In
the case of the scattering of particles with spin or of
many-particle processes the expansion of the total amplitude
is carried out in terms of matrix elements of the rotation group,
$d^j_{\mu\nu}(z)$~\cite{jw, w}, which satisfy the equation
\begin{equation}
\left[(1-z^2)\, \frac{d^{\,2}}{dz^2} - 2z\,\frac{d}{dz} +j(j+1)-
\frac{\mu^2-2\mu\nu z +\nu^2}{1-z^2}\, \right]\, y(z)=0 \,.
\label{eq}
\end{equation}
Therefore in order to continue the partial amplitudes of such
processes to complex values of the angular momentum it is first
necessary to generalize the functions $d^j_{\mu\nu}(z)$ and the
corresponding functions of the second kind to arbitrary values
of the indices and the argument. In the treatment of the elastic
scattering of particles with spin it is sufficient to have the
functions $d^j_{\mu\nu}(z)$ for arbitrary $j$ and $z$, but for
``physical'' values of the helicities $\mu$ and $\nu$ (i.e., for
$\mu$ and $\nu$ either both integers or both half-integers).
This case has been investigated, for example, in~\cite{hoa}. For
many-particle amplitudes, however, even a preliminary study
shows that they must obviously be continued not only with respect
to the total angular momentum, but also with respect to the
helicities~\cite{GPT}. Owing to this we shall consider in
the present paper the continuation of the functions $d^j_{\mu\nu}(z)$
to arbitrary values of $z,\,j,\,\mu,\,\nu\,$. Since, however, it
turns out that $d^j_{\mu\nu}(z)$ themselves have nonessential cuts
in  $j,\,\mu,$ and $\nu\,$, we study instead of them functions
which differ from $d^j_{\mu\nu}$ by a factor which does not depend
on $z\,$. These functions are a generalization of the associated
Legendre functions, so that it is natural to call them generalized
Legendre functions.

The present paper contains a description of the fundamental
properties of the generalized Legendre functions, and is an aid
in the study of many-particle amplitudes for complex angular
momenta.

\section{Definition of the generalized Legendre functions}

We define the Legendre functions of the first kind, $P^j_{\mu\nu}(z)$,
and of the second kind, $Q^j_{\mu\nu}(z)$, for arbitrary values of
$j,\,\mu,\,\nu,$ and $z$, by the equations
$$
P^j_{\mu\nu}(z)=
\frac1{\Gamma(\nu-\mu+1)}\,
\left(\frac{z-1}2\right)^{(\nu-\mu)/2}
\left(\frac{z+1}2\right)^{(\nu+\mu)/2}
~~~~~~~~~~~~~~~~~~~~~~~~~~~$$
\begin{equation}
~~~~~~~~~~~~\times F\left(j+\nu+1,-j+\nu;\,\nu-\mu+1;\frac{1-z}2\right)\,,
\label{defP}
\end{equation}
$$
Q^j_{\mu\nu}(z)= e^{i\pi(\mu-\nu)}\,
\frac{\Gamma(j+\mu+1)\Gamma(j-\nu+1)}{2\Gamma(2j+2)}\,
\left(\frac{z-1}2\right)^{-(j+1)}
\left(\frac{z+1}{z-1}\right)^{(\nu+\mu)/2}
$$
\begin{equation}
~~~~~~~~~~~~\times F\left(j+\nu+1,j+\mu+1;\,2j+2;\frac2{1-z}\right)\,.
\label{defQ}
\end{equation}
Here $F(a,b;c;x)$ is the hypergeometric function, and
$\mathrm{arg}(z-1)=\mathrm{arg}(z+1)=0$ for real $z>1$.
It is not hard to verify that the functions $P^j_{\mu\nu}(z)$
and $Q^j_{\mu\nu}(z)$ defined by (\ref{defP}) and (\ref{defQ})
satisfy Eq. (\ref{eq}).

It follows directly from (\ref{defP}) and (\ref{defQ}) that
for $\nu=0$ the functions $P^j_{\mu\nu}(z)$ and $Q^j_{\mu\nu}(z)$
go over into the associated Legendre functions:
\begin{equation}
P^j_{\mu 0}(z)=P^{\mu}_j(z)\,,~~~~~~
Q^j_{\mu 0}(z)=Q^{\mu}_j(z)\,,
\label{assoc}
\end{equation}
where $P^{\mu}_j(z)$ and $Q^{\mu}_j(z)$ are the associated Legendre
functions~\cite{Bat,GR}. In the general case the generalized
Legendre functions are connected in a simple way with the Jacobi
functions; for example,
$$P^j_{\mu\nu}(z)=\frac{\Gamma(j-\nu+1)}{\Gamma(j-\mu+1)}\,
\left(\frac{z-1}2\right)^{(\nu-\mu)/2}
\left(\frac{z+1}2\right)^{(\nu+\mu)/2}
P_{j-\nu}^{\nu-\mu,\,\nu+\mu}(z)\,, ~~~~~~~~~~~~~~ $$
\begin{equation}
Q^j_{\mu\nu}(z)= e^{i\pi(\mu-\nu)}\,
\frac{\Gamma(j-\nu+1)}{\Gamma(j-\mu+1)}\,
\left(\frac{z-1}2\right)^{(\nu-\mu)/2}
\left(\frac{z+1}2\right)^{(\nu+\mu)/2}
Q_{j-\nu}^{\nu-\mu,\,\nu+\mu}(z)\,,
\label{jac}
\end{equation}
where $P_n^{\alpha,\beta}(z)$ and $Q_n^{\alpha,\beta}(z)$
are the Jacobi functions\footnote{Note that $P_n^{\alpha,\beta}(z)$
with positive integer $n$ is the Jacobi polynomial of order $n$.}
 of the first and second kinds~\cite{Bat}.

We shall list some simple properties of the functions $P^j_{\mu\nu}$
and $Q^j_{\mu\nu}$ which follow directly from the definitions
(\ref{defP}) and (\ref{defQ}). It is obvious from (\ref{defP}) that
\begin{equation}
P^j_{\mu\nu}(z)=P^{-j-1}_{\mu\nu}(z)\,.
\label{P-j}
\end{equation}

It is clear from the definition (\ref{defQ}) that $Q^j_{\mu\nu}$
and $Q^j_{\nu\nu}$ differ only by a factor:
\begin{equation}
Q^j_{\mu\nu}(z)= e^{2i\pi(\mu-\nu)}\,\,
\frac{\Gamma(j+\mu+1)\,\Gamma(j-\nu+1)}
{\Gamma(j-\mu+1)\,\Gamma(j+\nu+1)}\,\,Q^j_{\nu\mu}(z)\,.
\label{Qmn}
\end{equation}
The functions $P^j_{\mu\nu}$ and $P^j_{\nu\nu}\,$, on the other
hand, are in general linearly independent solutions of Eq.(\ref{eq}),
with the Wronskian determinant
$$
W(P^j_{\mu\nu},P^{j}_{\nu\mu})=\frac2{\pi}\,
\frac{\sin\pi(\nu-\mu)}{1-z^2}\,,
$$
where $W(f_1,\,f_2)=f_1\,f'_2-f'_1\,f_2\,$. The functions
$Q^j_{\mu\nu}$ and $Q ^{-j-1}_{\nu\mu}$ are likewise linearly
independent. Their Wronskian determinant is
$$
W(Q^j_{\mu\nu},Q^{-j-1}_{\nu\mu})=\frac{\pi}2\,\,
\frac{\sin2\pi j}{\sin\pi(j+\mu)\,\sin\pi(j-\nu)}\,\,
\frac1{1-z^2}\,.
$$

Using the properties of hypergeometric functions, one can also derive
other relations for the generalized Legendre functions. The simplest
of these are the equations
\begin{equation}
P^j_{\mu\nu}(z)=P^{j}_{-\nu,-\mu}(z)\,,~~~~~~
Q^j_{\mu\nu}(z)=Q^{j}_{-\nu,-\mu}(z)\,,
\label{munu}
\end{equation}
which follow from Eq. 2.1(23) in~\cite{Bat} (9.131.1 in~\cite{GR}).
From Eq. 2.1(2.2) in~\cite{Bat} one also gets another definition
for $P^j_{\mu\nu}(z)$, which is equivalent to the definition
(\ref{defP}):
$$
P^j_{\mu\nu}(z)=
\frac1{\Gamma(\nu-\mu+1)}\,
\left(\frac{z+1}2\right)^j\,
\left(\frac{z-1}{z+1}\right)^{(\nu-\mu)/2}\,
~~~~~~~~~~~~~~~~~~~~~~~~~~~~~~$$
$$~~~~~~~~~~~~~~~~~~~~~~~~~~~~~~~~~~~~~~~~~~~~\times
F\left(-j+\nu,-j-\mu;\,\nu-\mu+1;\,\frac{z-1}{z+1}\right)\,.
~~~~~~~~~~~~~~~~~~~~~~~~\textrm{(2a)}
$$
The form (2a) for $P^j_{\mu\nu}$ makes the first of the equations
(\ref{munu}) obvious.

\section{Recurrence relations}

As is well known, adjacent hypergeometric functions are connected
by linear relations. These lead to the following recurrence
relations for the generalized Legendre functions:
\begin{equation}
(2j+1)\,\sqrt{\frac{z-1}2}\,\,P^j_{\mu-1/2,\,\nu+1/2}(z)=
P^{j+1/2}_{\mu\nu}(z)-P^{j-1/2}_{\mu\nu}(z)\,,
~~~~~~~~~~~~~~~~~~
\label{-+}
\end{equation}
$$
(2j+1)\,\sqrt{\frac{z-1}2}\,\,P^j_{\mu+1/2,\,\nu-1/2}(z)=
(j+\nu+1/2)(j-\mu+1/2)\,P^{j+1/2}_{\mu\nu}(z)
$$
\begin{equation}
~~~~~~~~~~~~~~~ -(j-\nu+1/2)(j+\mu+1/2)\, P^{j-1/2}_{\mu\nu}(z)\,,
\label{+-}
\end{equation}
$$
(2j+1)\,\sqrt{\frac{z+1}2}\,\,P^j_{\mu+1/2,\,\nu+1/2}(z)=
(j-\mu+1/2)\,P^{j+1/2}_{\mu\nu}(z)~~~~~~~~~~~~~~~
$$
\begin{equation}
+(j+\mu+1/2)\,P^{j-1/2}_{\mu\nu}(z)\,,
\label{++}
\end{equation}
$$
(2j+1)\,\sqrt{\frac{z+1}2}\,\,P^j_{\mu-1/2,\,\nu-1/2}(z)=
(j+\nu+1/2)\,P^{j+1/2}_{\mu\nu}(z)~~~~~~~~~~~~~~
$$
\begin{equation}
+(j-\nu+1/2)\,P^{j-1/2}_{\mu\nu}(z)\,,
\label{--}
\end{equation}
Equations (\ref{-+}) and (\ref{+-}) follow from the respective
formulas 9.137.4 and 9.137.5 in~\cite{GR}, and (\ref{++})
and (\ref{--}) follow from Eqs. 2.8(37) and 2.8(32) in~\cite{Bat}.

By iteration of the relations (\ref{-+})--(\ref{--}) we can derive
a large number of other recurrence relations. One example of these is
$$
j(j+1)(2j+1)\,z\,P^j_{\mu\nu}(z)=j(j+\nu+1)(j-\mu+1)\,P^{j+1}_{\mu\nu}(z)
~~~~~~~~~~$$
\begin{equation}
~~~~~~~~~~~~~~~~~~ +\nu\mu\,(2j+1)\,P^j_{\mu\nu}(z)
+(j+1)(j+\mu)(j-\nu)\,P^{j-1}_{\mu\nu}(z)\,.
\label{z}
\end{equation}

Another type of recurrence relations can be derived by using the
differentiation properties of hypergeometric functions [Eqs. 2.8(27)
and 2.8(20) in~\cite{Bat}]. We thus get the following equations:
$$
\frac{d^n}{dz^n} [(z-1)^{(\nu-\mu)/2}(z+1)^{(\nu+
\mu)/2}\,P^j_{\mu\nu}(z)] ~~~~~~~~~~~~~~
$$
\begin{equation}
~~~~~~~~~~~~~~~~~~~~~~~~~~~~~~~~~~~~ =(z-1)^{(\nu-\mu-n)/2}(z+1)^{(\nu+
\mu-n)/2}\,P^j_{\mu,\nu-n}(z)\,,
\label{d-n}
\end{equation}
$$
\frac{\Gamma(j+\nu+1)}{\Gamma(j-\nu+1)}\,
\frac{d^n}{dz^n} [(z-1)^{-(\nu-\mu)/2}(z+1)^{-(\nu+
\mu)/2}\,P^j_{\mu\nu}(z)]~~~~~~~~~~~~~~~~~~~~~~~~~~
$$
\begin{equation}
~~~~~~~~~~~~~~~~~ =\frac{\Gamma(j+\nu+n+1)}{\Gamma(j-\nu-n+1)}\,
(z-1)^{-(\nu-\mu+n)/2}(z+1)^{-(\nu+\mu+n)/2}\,P^j_{\mu,\nu+n}(z)\,.
\label{d+n}
\end{equation}

Two further formulas of this type, giving changes of the index $\mu$,
are obtained if we make the replacement $\mu \leftrightarrow -\nu$
in (\ref{d-n}) and (\ref{d+n}) and use the property (\ref{munu}).

All of the formulas (\ref{-+})--(\ref{d+n}) are valid not only for
$P^j_{\mu\nu}$, but also for the functions of the second kind,
$Q^j_{\mu\nu}$.

\section{Relations between the Functions  of 1st
and 2nd Kinds}

Equation (\ref{eq}) has two linearly independent solutions.
Therefore it is clear that there must exist relations between
the four different solutions of this equation: $P^j_{\mu\nu}(z),\,\,
P^j_{\nu\mu}(z),\,\, Q^j_{\mu\nu}(z),$ and $Q^{-j-1}_{\mu\nu}(z)$.
We can derive these relations easily by using Eq. 2.10(2) of~\cite{Bat}
(or Eq. 9.132.2 of~\cite{GR}). Applying this equation to (\ref{defQ}),
we have
\begin{equation}
\frac2{\pi}\,e^{-i\pi(\mu-\nu)}\,\sin\pi(\mu-\nu)\, Q^j_{\mu\nu}(z)
=P^j_{\mu\nu}(z)
-\frac{\Gamma(j+\mu+1)\,\Gamma(j-\nu+1)}
{\Gamma(j-\mu+1)\,\Gamma(j+\nu+1)}\,P^j_{\nu\mu}(z)\,.
\label{QPP}
\end{equation}

Recalling the property (\ref{P-j}), we easily define a further equation
$$
Q^j_{\mu\nu}(z)-Q^{-j-1}_{\mu\nu}(z)~~~~~~~~~~~~~~~~~~~~~~~~~~~~~~~
~~~~~~~~~~~~~~~~~~~~~~~~~~~~~~~~~~~~~
$$
\begin{equation}
~~~~~~~~~~~~~ =\frac{\pi}2\,e^{i\pi(\mu-\nu)}\,\frac{\sin2\pi j}
{\sin\pi(j-\mu)\sin\pi(j+\nu)}\,
\frac{\Gamma(j+\mu+1)\,\Gamma(j-\nu+1)}
{\Gamma(j-\mu+1)\,\Gamma(j+\nu+1)}\,P^j_{\nu\mu}(z)\,.
\label{QQP}
\end{equation}

It is furthermore obvious that Eq. (\ref{eq}) remains unchanged if we
change the sign of the variable $z$ and at the same time change the
sign of one of the indices, $\mu$ or $\nu$. Therefore there are also
relations connecting generalized Legendre functions of $z$ and of $-z$.
The simplest of these is obtained by applying to Eq. (\ref{defQ}) the
formula 2.10(6) from~\cite{Bat} (or 9.131.1 from~\cite{GR}) for the
hypergeometric function:
$$
Q^j_{\mu\nu}(z)= e^{\mp i(j+1)\pi}\,e^{-2i\pi\nu}\,
\frac{\Gamma(j-\nu+1)}{\Gamma(j+\nu+1)}\,Q^j_{\mu,-\nu}(-z)~~~~~~
$$
\begin{equation}
~~~~~~ =e^{\mp i(j+1)\pi}\,e^{2i\pi\mu}\,
\frac{\Gamma(j+\mu+1)}{\Gamma(j-\mu+1)}\,Q^j_{-\mu,\nu}(-z)\,.
\label{Q-z}
\end{equation}
The sign $\mp$ corresponds to values $\mathrm{Im}z\gtrless 0\,$.

When we now use (\ref{QQP}) and (\ref{Q-z}), it is not hard to
derive the following relations:
$$
\frac{\Gamma(j+\nu+1)}{\Gamma(j-\nu+1)}\,P^j_{\mu\nu}(z)
=e^{\pm i\pi j}\,P^j_{\mu,-\nu}(-z)~~~~~~~~~~~~~~~~~~~~
$$
\begin{equation}
~~~~~~~~~~~~~~~~~~~ -\frac2{\pi}\,
e^{\pm i\pi\nu}\,e^{-i\pi(\mu+\nu)}\,\sin\pi(j+\mu)\,
Q^j_{\mu,-\nu}(-z)\,,
\label{Pmu-z}
\end{equation}
$$
\frac{\Gamma(j-\mu+1)}{\Gamma(j+\mu+1)}\,P^j_{\mu\nu}(z)
=e^{\pm i\pi j}\,P^j_{-\mu,\nu}(-z)~~~~~~~~~~~~~~~~~~~~
$$
\begin{equation}
~~~~~~~~~~~~~~~~~~~ -\frac2{\pi}\,
e^{\mp i\pi\mu}\,e^{i\pi(\mu+\nu)}\,\sin\pi(j-\nu)\,
Q^j_{-\mu,\nu}(-z)\,,
\label{Pnu-z}
\end{equation}
$$
~~~\frac1{\pi}\,\sin\pi(\mu+\nu)\,P^j_{\mu\nu}(z)
=\frac{e^{\mp i\pi\mu}}{\Gamma(j+\nu+1)\,\Gamma(-j+\nu)}\,
 P^j_{\mu,-\nu}(-z)
$$
\begin{equation}
~~~~~~~~~~~~~~~~~~~~ -\frac{e^{\pm i\pi\nu}}{\Gamma(j-\mu+1)\,
\Gamma(-j-\mu)}\,
P^j_{-\mu,\nu}(-z)\,.
\label{PP-z}
\end{equation}
In Eqs. (\ref{Pmu-z})--(\ref{PP-z}), as in (\ref{Q-z}),
the upper signs correspond to the case $\mathrm{Im}z>0$, and
the lower signs to $\mathrm{Im}z<0$~\footnote{Recall that the
definitions (\ref{defP}) and (\ref{defQ}) unambiguously fix the
phases of the Legendre functions at real $z>+1$. Relations
(\ref{Q-z})--(\ref{PP-z}) define analytical continuation
of these functions to real $z<-1$ through upper or lower
complex half-plane.}.

\section{Analytic Properties}

The analytic properties of the generalized Legendre functions are
clear from their definitions (\ref{defP}), (\ref{defQ}). The function
of the first kind, $P^j_{\mu\nu}(z)$, is an entire function of each
of the three indices $j,\,\mu,$ and $\nu$. It has zeros,
$$
P^j_{j+n+1,\,j-m}(z)\equiv 0\,,~~~~~
P^j_{-j+m,\,-j-n-1}(z)\equiv 0\,,
$$
if $m\ge 0$ and $n\ge 0$ are nonnegative integers.

The function of the second kind, $Q^j_{\mu\nu}(z)$, is a meromorphic
function of the indices. It has poles at $j+\mu+1=-n$ or $j-\nu+1=-n$,
where $n$ is a nonnegative integer. The residues at the poles of the
function $Q^j_{\mu\nu}(z)$ can be expressed in terms of $Q^j_{\nu\mu}(z)$
or $P^j_{\nu\mu}(z)$ by means of Eqs. (\ref{Qmn}) and (\ref{QPP}).
Coincidence of two poles of $Q^j_{\mu\nu}(z)$ in the $j$ plane in general
leads to a pole of second order with respect to $j$. For $\nu=0$ there
is no second order pole.

As functions of the variable $z$ the generalized Legendre functions are
analytic in the complex plane with two cuts drawn along the real axis
from $-\infty$ to $-1$ and from $-1$ to $+1$. The discontinuities on the
cut that goes from $-\infty$ to $-1$ can be calculated easily from
(\ref{Q-z})--(\ref{PP-z}):
$$
\frac1{2i}\,[Q^j_{\mu\nu}(x+i\epsilon)-Q^j_{\mu\nu}(x-i\epsilon)]
=e^{-2i\pi\nu}\,\sin{\pi j}\,
\frac{\Gamma(j-\nu+1)}{\Gamma(j+\nu+1)}\,
Q^j_{\mu,-\nu}(-x)\,,
$$
\begin{equation}
~~~~~~~~~~~~~~~~~~~~~~~~~~~~~~~~~~~~~~~~~~
~~~~~~~~~~~~~~~~~~~~ -\infty<x<-1\,,
\label{Qlcut}
\end{equation}
$$
\frac1{2i}\,[P^j_{\mu\nu}(x+i\epsilon)-P^j_{\mu\nu}(x-i\epsilon)]
=\frac{\Gamma(j-\nu+1)}{\Gamma(j+\nu+1)}\,
[\,\sin\pi j\,P^j_{\mu,-\nu}(-x)~~~~~~~~~
$$
\begin{equation}
~~~~~ -\frac2{\pi}\,e^{-i\pi(\mu+\nu)}\,\sin\pi\nu\,\sin\pi(j+\mu)\,
Q^j_{\mu,-\nu}(-x)\,]\,,
~~~~~ -\infty<x<-1\,.
\label{Plcut}
\end{equation}
Other expressions for these discontinuities are obtained by interchanging
$\mu\leftrightarrow-\nu$ and using the properties (\ref{munu}).

To study the cut that goes from $-1$ to $+1$ it is convenient to introduce
the function $\widetilde{P}^j_{\mu\nu}(x)$ defined by the equations
\begin{equation}
\widetilde{P}^j_{\mu\nu}(x)
=e^{i\pi(\mu-\nu)/2}\,
P^j_{\mu\nu}(x+i\epsilon)
=e^{-i\pi(\mu-\nu)/2}\,
P^j_{\mu\nu}(x-i\epsilon)\,,~~ -1<x<+1\,.
\label{defPwt}
\end{equation}
We then have the obvious expression
\begin{equation}
\frac1{2i}\,[P^j_{\mu\nu}(x+i\epsilon)-P^j_{\mu\nu}(x-i\epsilon)]
=\sin\frac{\pi}2(\nu-\mu)\,\widetilde{P}^j_{\mu\nu}(x)\,,
~~~~~~~ -1<x<+1\,.
\label{Prcut}
\end{equation}

The discontinuity of $Q^j_{\mu\nu}(z)$  can be expressed in terms
of $\widetilde{P}^j_{\mu\nu}$ and $\widetilde{P}^j_{\nu\mu}$ by
means of (\ref{QPP}) and (\ref{Prcut}). A more important formula,
however, is
$$
\frac1{2i}\,[e^{i\pi(\nu-\mu)/2}\,Q^j_{\mu\nu}(x+i\epsilon)
-e^{-i\pi(\nu-\mu)/2}\,Q^j_{\mu\nu}(x-i\epsilon)]~~~~~~~~~~~~~~
$$
\begin{equation}
~~~~~~~~~~~~~~~~~~~~~~~~~~~~ =-\frac{\pi}2\,e^{-i\pi(\nu-\mu)}\,
\widetilde{P}^j_{\mu\nu}(x)\,,
~~~~~~~~~~~~~~~~~~~ -1<x<+1\,.
\label{Qrcut}
\end{equation}

\section{The Asymptotic Behavior}

The definitions (\ref{defP}) and (\ref{defQ}) allow us to study
with ease the asymptotic behavior of the generalized Legendre
functions with respect to the variable $z$. For example, (\ref{defP})
describes the behavior of the function of the first kind,
$P^j_{\mu\nu}(z)$, for $(z-1)\to0$. The behavior of the functions
of the second kind is then found from the expression (\ref{QPP}).
The asymptotic behavior for $z\to\infty$ is given by (\ref{defQ})
and (\ref{QQP}). The behavior of the functions $P^j_{\mu\nu}$ and
$Q^j_{\mu\nu}$ for $(z+1)\to0$ can be found without difficulty
by means of (\ref{PP-z}).

Let us now proceed to consider the asymptotic behavior of the
generalized Legendre functions with respect to the indices. For this
it is convenient to rewrite (\ref{eq}) in a different form, using
the change of variable $z=\cosh\alpha$
$$
\left[\frac{d^2}{d\alpha^2}+\coth\alpha\,\frac{d}{d\alpha}
-\frac1{\sinh^2(\alpha/2)}\,\left(\frac{\nu-\mu}2\right)^2
+\frac1{\cosh^2(\alpha/2)}\,\left(\frac{\nu+\mu}2\right)^2
\right]\,y(\cosh\alpha)~~~~
$$
$$
~~~~~~~~~~~~~~~~~~~~~~~~~~~~~~~~~~~~~~
~~~~~~~~~~~~~~~~~~~~~~~~~~~~~ =j(j+1)\,y(\cosh\alpha)\,.
~~~~~~~~~~~~(1a)
$$

With this way of rewriting (\ref{eq}) and the asymptotic form
of $P^j_{\mu\nu}(z)$ for $z\to1$, we can get the following
value of a limit:
\begin{equation}
\lim_{t\to\infty} \left[\,t^{(\nu-\mu)/2}\,
P^j_{\mu\nu}\left(\cosh\frac{y}{\sqrt{t}}\right)\right]
=J_{\nu-\mu}(y)\,,
\label{PJ}
\end{equation}
where $J_\kappa(y)$ is the Bessel function and $t=(\nu+\mu)^2/4-
j(j+1)$; it is supposed that $\nu-\mu=\mathrm{const}$ and
$(\nu+\mu)/t\to0$ as $t\to\infty$.

In the more general case we have
$$
\lim_{t\to\infty} \left[\,t^{(\nu-\mu)/2}\,
P^j_{\mu\nu}\left(1+\frac{2x}{t}\right)\right]~~~~~~~~~~~~~~~~~~
~~~~~~~~~~~~~~~~~~~
$$
\begin{equation}
~~~~~~~~~~~ =\frac{x^{(\nu-\mu)/2}}{\Gamma(\nu-\mu+1)}\,\,e^{ax/2}\,\,
\Phi\left(\frac{\nu-\mu+1}2+\frac{b}{a};\,\nu-\mu+1;\,-ax\right)\,,
\label{PPhi}
\end{equation}
where
$$
a=\lim_{t\to\infty} (\nu+\mu)/t\,,~~~~
b=\lim_{t\to\infty} [(\nu+\mu)^2/4-j(j+1)]/t\,,
$$
and $\Phi$ is the confluent hypergeometric
function~\cite{Bat,GR}~\footnote{In this limit, the limiting expressions
for the generalized Legendre functions of the 1st and 2nd kinds may be
expressed also through the Whittaker functions~\cite{Bat,GR}, of the 1st
and 2nd kinds respectively.}. It is not hard to verify that for
$a\to0,\,\,b\to1$ the relation (\ref{PPhi}) goes into (\ref{PJ}).
Equations (\ref{PJ}) and (\ref{PPhi}) can be used as an approximate
expression for $P^j_{\mu\nu}(z)$ when $z$ is very close to unity.

The asymptotic expansion for $\mu\to\infty$ and fixed $z,\,j,\,\nu$
is obtained from the definition (\ref{defP}). To study other cases of
the asymptotic behavior of generalized Legendre functions with respect
to the indices for fixed $z$ it is convenient to apply to Eq. (1a)
a method analogous to the quasiclassical approximation of Wentzel, Kramers,
and Brillouin, and use the well known behavior of the solutions with
respect to $z$. In this way it is not hard to find that with $\mu$
and $\nu$ constant
\begin{equation}
Q^j_{\mu\nu}(\cosh\alpha)_{j\to \infty} \sim e^{i\pi(\mu-\nu)}\,
j^{\mu-\nu-1/2}\,\sqrt{\frac{\pi}{2\sinh\alpha}}\,\,
e^{-\alpha(j+1/2)}\,.
\label{asQ}
\end{equation}

The following asymptotic cases can be treated similarly:

1) for $j,\,\mu+\nu = \mathrm{const}$
\begin{equation}
P^j_{\mu\nu}(\cosh\alpha)_{\nu-\mu\to \infty} \sim \frac1{\Gamma(\nu-\mu+1)}\,
\frac{[2\tanh(\alpha/4)]^{\nu-\mu}}
{\sqrt{\cosh(\alpha/2)}}\,,~~~~
\label{asP}
\end{equation}

2) for $\nu,\, j-\mu = \mathrm{const}$, and $j,\, \mu \to\infty$
$$
Q^j_{\mu\nu}(\cosh\alpha) \sim e^{i\pi(\mu-\nu)}\,
\frac{\Gamma(j+\mu+1)}{\Gamma(j+\nu+1)}\,
\sqrt{\frac{2\pi}{j+\mu+1}}~~~~~~~~~~
$$
$$
~~~~~~~~~~\times[\tanh(\alpha/2)]^\nu\,
(\tanh\alpha)^{-\mu}
\,(2\cosh\alpha)^{-(j+1)}\,,
$$
$$
Q^{-j-1}_{\mu\nu}(\cosh\alpha) \sim e^{i\pi(\mu-\nu)}\,
\frac{\Gamma(-j+\mu)}{\Gamma(-j+\nu)}\,
\sqrt{\frac{2\pi}{-j-\mu}}~~~~~~~~~~~~
$$
\begin{equation}
~~~~~~~~\times[\tanh(\alpha/2)]^{-\nu}\,
(\tanh\alpha)^{\mu}
\,(2\cosh\alpha)^{j}\,,
\label{asQj}
\end{equation}

The asymptotic formulas (\ref{PJ})--(\ref{asQj}) are written out for
the functions
for which they have the simplest forms. The asymptotic behaviors
of other functions in these cases can be found by using the
relations (\ref{P-j}), (\ref{Qmn}), (\ref{QPP}), (\ref{QQP}).

\section{Integrals of Products of two Functions. Orthogonal Systems}

Equation (\ref{eq}) allows us to calculate easily the indefinite
integral of the product of $f_{j_1}$ and $f_{j_2}$, where $f_{j_1}$
and $f_{j_2}$ are solutions of (\ref{eq}) with the same $\mu,\,\nu$
but different values of $j$
\begin{equation}
\int f_{j_1}(z)\, f_{j_2}(z)\,dz=
\frac{1-z^2}{(j_1-j_2)(j_1+j_2+1)}\,
(f_{j_2}\, f'_{j_1}-f'_{j_2}\, f_{j_1})\,.
\label{int}
\end{equation}
A simple example is the integral
\begin{equation}
\int_1^\infty dz\,P^j_{\mu\nu}(z)\, Q^l_{\nu\mu}(z)=
\frac{e^{i\pi(\nu-\mu)}}{(l-j)(l+j+1)}\,.
\label{iPQ}
\end{equation}
Convergence of the integral (\ref{iPQ}) at the upper limit imposes
the requirement \mbox{$\mathrm{Re}\,l>\mathrm{Re}j\ge -1/2\,$,} and
convergence at the lower limit requires $\mathrm{Re}(\nu-\mu+1)>0\,$.

In precisely the same way we can calculate the integral from
$-1$ to $+1$ of the product $\widetilde{P}^{j_1}_{\mu\nu}(x)
\widetilde{P}^{j_2}_{\mu\nu}(x)\,$. In general the existence
of this integral requires some other conditions besides
$\mathrm{Re}(\nu-\mu+1)>0\,$. Both the result and the conditions
in general form are rather cumbersome, and we shall not write
them out here. It is essential, however, to point out that the
functions $\widetilde{P}^{j}_{\mu\nu}(x)$ contain two systems
orthogonal on the interval $[-1,\,1]\,$. One of these systems
is determined by the requirements
\begin{equation}
\mathrm{Re}(\nu-\mu+1)>0\,,~~~~
\mathrm{Re}(\nu+\mu+1)>0\,,~~~~
j-\nu=n\ge0\,,
\label{cond}
\end{equation}
where $n$ is an integer. The other system is determined by the
requirements that are obtained from (\ref{cond}) by the interchange
$\mu\leftrightarrow -\nu\,$. As the first of the equations (\ref{jac})
shows, both orthogonal systems are connected with the Jacobi
polynomials. If the function $\widetilde{P}^{j}_{\mu\nu}(x)$
belongs to one of the orthogonal systems, its norm is
\begin{equation}
\int_{-1}^{+1} dx\,[\widetilde{P}^j_{\mu\nu}(x)]^2=
\frac2{2j+1}\,\frac{\Gamma(j+\mu+1)\,\Gamma(j-\nu+1)}
{\Gamma(j-\mu+1)\,\Gamma(j+\nu+1)}\,.
\label{iPP}
\end{equation}

As an example of the expansions that can occur we can present
the following series:
$$
\frac1{\zeta-z}= e^{i\pi(\mu-\nu)}\,
\left(\frac{z-1}{\zeta-1}\right)^{-(\nu-\mu)/2}\,
\left(\frac{z+1}{\zeta+1}\right)^{-(\nu+\mu)/2}~~~~~~~~~~
$$
\begin{equation}
~~~~~~~~~~~\times\sum_{j-\nu=n=0}^{\infty} (2j+1)\,
P^j_{\mu\nu}(z)\,Q^j_{\nu\mu}(\zeta)\,.
\label{ser}
\end{equation}

\section{The Functions  $d^j_{\mu\nu}(x)$ }

Let us now establish the connection between the generalized
Legendre functions and the functions $d^j_{\mu\nu}(x)\,$.
If all of the indices $j,\,\mu,$ and $\nu$ are either integers,
or else all half-integers, and if $-|j+1/2|<(\mu,\,\nu)<|j+1/2|\,$,
then we can define
\begin{equation}
d^j_{\mu\nu}(x)=\widetilde{P}^j_{\mu\nu}(x)\,
\left[\frac{\Gamma(j-\mu+1)\,\Gamma(j+\nu+1)}
{\Gamma(j+\mu+1)\,\Gamma(j-\nu+1)}\right]^{1/2}\,.
\label{d}
\end{equation}

By the use of (\ref{P-j}), (\ref{munu}), (\ref{PP-z}) and (\ref{defPwt})
it is easy to show that the following equations hold for $d^j_{\mu\nu}(x)$:
$$
d^j_{\mu\nu}(x)=d^{-j-1}_{\mu\nu}(x)=d^j_{-\nu,-\mu}(x)
=(-1)^{\mu-\nu}d^j_{\nu\mu}(x)~~~~
$$
\begin{equation}
~~~~~~~~ =(-1)^{j-\nu}d^j_{-\mu,\nu}(-x)
=(-1)^{j+\mu}d^j_{\mu,-\nu}(-x)\,.
\label{d-x}
\end{equation}
Besides this, the functions $d^j_{\mu\nu}(x)$ with different $j$
are orthogonal, and it follows from (\ref{iPP}) that
\begin{equation}
\int_{-1}^{+1} d^j_{\mu\nu}(x)\,d^l_{\mu\nu}(x)\,dx
=\frac2{2j+1}\,\delta_{jl}\,.
\label{idd}
\end{equation}
It is not hard to verify that the definition (\ref{d}) corresponds
to the choice of the functions $d^j_{\mu\nu}$ that is
commonly used (cf., e.g.,~\cite{jw}).

\section{Addition and Multiplication Formulas}

Addition theorems play a large part in the theory of Legendre
functions, and also in that of the functions $d^j_{\mu\nu}\,$.
It turns out that also for the generalized Legendre functions
one can establish addition theorems in general form.

We first establish the equation
\begin{equation}
Q^j_{\mu\lambda}(z_1)\,Q^j_{\lambda\mu}(z_2)
=\frac12\,\int_{-\infty}^{\infty}d\alpha\,e^{-\lambda\alpha}\,
e^{\mu\alpha_2}\,Q^j_{\mu\nu}(z_\alpha)\,e^{\nu\alpha_1}\,;
\label{Qa}
\end{equation}
here $z_\alpha,\,\alpha_1,$ and $\alpha_2$ are defined by
the relations\footnote{We assume (at least, initially) that
$z_1,\,z_2>+1$; then the sigh before the last square root
coincides with the sign of $\alpha$ (the same is true for signs
of $\alpha_1,\,\alpha_2$).}
$$
z_\alpha=z_1\,z_2+\sqrt{z_1^2-1}\,\sqrt{z_2^2-1}\,\cosh\alpha\,,~
$$
$$
z_1=z_2\,z_\alpha-\sqrt{z_2^2-1}\,\sqrt{z_\alpha^2-1}\,\cosh\alpha_1\,,
$$
$$
z_2=z_\alpha\,z_1-\sqrt{z_\alpha^2-1}\,\sqrt{z_1^2-1}\,\cosh\alpha_2\,,
$$
$$
\sinh\alpha\,\sqrt{z_1^2-1}\,\sqrt{z_2^2-1}
=\sinh\alpha_1\,\sqrt{z_2^2-1}\,\sqrt{z_\alpha^2-1}
=\sinh\alpha_2\,\sqrt{z_\alpha^2-1}\,\sqrt{z_1^2-1}
$$
\begin{equation}
=\pm\sqrt{z_\alpha^2+z_1^2+z_2^2-2z_\alpha\,z_1\,z_2-1}\,.
\label{z's}
\end{equation}
The requirement that the integral in (\ref{Qa}) converges imposes
the restrictions
$$
\mathrm{Re}(j-\lambda+1)>0\,,~~~~\mathrm{Re}(j+\lambda+1)>0\,.
$$

To prove Eq. (\ref{Qa}), it suffices to note the following. The
right member of (\ref{Qa}) satisfies the same equation with respect
to $z_1$ and $z_2$ as the left member. (This can be verified by a
direct but rather cumbersome calculation.) When $z_1$ or $z_2$ increases,
there is an equally rapid decrease of the right and left members.
For $z_1\to\infty$ and $z_2\to\infty$ the coefficient in the asymptotic
form is the same for the right and left members.

From (\ref{Qa}) one easily gets the inverse equation
\begin{equation}
e^{\mu\alpha_2}\,Q^j_{\mu\nu}(z_\alpha)\,e^{\nu\alpha_1}
=\frac1{\pi i}\,\int_{-i\infty}^{i\infty}d\lambda\,
e^{\lambda\alpha}\,Q^j_{\mu\lambda}(z_1)\,Q^j_{\lambda\nu}(z_2)\,.
\label{Qlam}
\end{equation}
In (\ref{Qlam}) it is assumed that $\mathrm{Re}j$ is sufficiently
large so that the poles of the integrand do not fall on the contour.
$\alpha$ is originally real, but (\ref{Qlam}) can be continued
analytically with respect to $\alpha_1,\,,z_1,$ and $z_2$ as long as
the integral does not become divergent at infinity. For further
continuation it is necessary to break up the integrand into
components and rotate the path of integration. The path is then
turned differently for different terms.

For $\mu=\nu=0$ (\ref{Qlam}) is an integral form of the usual
addition theorem for Legendre functions of the second kind [Eq. 3.11(4)
in~\cite{Bat} or 8.795.2 in~\cite{GR}]. In the general case it
determines the addition theorem for generalized Legendre functions.
Equation (\ref{Qa}) is the inverse of (\ref{Qlam}), and it is
natural to call it a multiplication formula.

By means of (\ref{Qa}) and (\ref{Qlam}) we can obtain a large
number of other addition and multiplication formulas.

For example, using (\ref{Qrcut}) and (\ref{defPwt}) we can show
that for $z_1>z_2$ and $\mathrm{Re}(j-\lambda+1)>0$
$$
e^{i\pi(\lambda-\nu)}\,Q^j_{\mu\lambda}(z_1)\,P^j_{\lambda\nu}(z_2)
=\frac1{2\pi}\int_{-\pi}^{\pi}d\theta\,e^{-i\lambda\theta}\,
e^{-i\mu\theta_2}\,Q^j_{\mu\nu}(z_\theta)\,e^{-i\nu\theta_1}
$$
\begin{equation}
~~~~~~~~~~~~+\frac1{\pi}\,\sin\pi(\lambda-\nu)\,
\int^0_{-\infty}d\alpha\,e^{-\lambda\alpha}\,
e^{\mu\alpha_2}\,Q^j_{\mu\nu}(z_\alpha)\,e^{\nu\alpha_1}\,,
\label{PQa-}
\end{equation}
and for $z_1<z_2$ and $\mathrm{Re}(j+\lambda+1)>0$
$$
e^{i\pi(\mu-\lambda)}\,P^j_{\mu\lambda}(z_1)\,Q^j_{\lambda\nu}(z_2)
=\frac1{2\pi}\int_{-\pi}^{\pi}d\theta\,e^{-i\lambda\theta}\,
e^{-i\mu\theta_2}\,Q^j_{\mu\nu}(z_\theta)\,e^{-i\nu\theta_1}
$$
\begin{equation}
~~~~~~~~~~~~+\frac1{\pi}\,\sin\pi(\mu-\lambda)\,
\int_0^{\infty}d\alpha\,e^{-\lambda\alpha}\,
e^{\mu\alpha_2}\,Q^j_{\mu\nu}(z_\alpha)\,e^{\nu\alpha_1}\,.
\label{PQa+}
\end{equation}
The quantities $z_\theta,\,\theta_1,$ and $\theta_2$ in (\ref{PQa-})
and (\ref{PQa+}) are defined by the relations\footnote{Again, we
assume $z_1,\,z_2>+1$; the sign before the square root (as well as
the signs of $\sin\theta_1$ and $\sin\theta_2$) coincides with the
sign of $\sin\theta$. }
$$
z_\theta=z_1\,z_2-\sqrt{z_1^2-1}\,\sqrt{z_2^2-1}\,\cos\theta\,,~~
$$
$$
z_1=z_2\,z_\theta-\sqrt{z_2^2-1}\,\sqrt{z_\theta^2-1}\,\cos\theta_1\,,
$$
$$
z_2=z_\theta\,z_1-\sqrt{z_\theta^2-1}\,\sqrt{z_1^2-1}\,\cos\theta_2\,,
$$
$$
\sin\theta\,\sqrt{z_1^2-1}\,\sqrt{z_2^2-1}
=\sin\theta_1\,\sqrt{z_2^2-1}\,\sqrt{z_\theta^2-1}
=\sin\theta_2\,\sqrt{z_\theta^2-1}\,\sqrt{z_1^2-1}
$$
\begin{equation}
=\pm\sqrt{1+2z_\theta\,z_1\,z_2-z_\theta^2-z_1^2-z_2^2}\,.
\label{z's}
\end{equation}

We note that in the limit $z_1\to\infty\,$ Eqs. (\ref{Qa}) and
(\ref{PQa-}) give integral representations for the generalized
Legendre functions, for example,
$$
Q^j_{\mu\nu}(z)=\frac12\,e^{i\pi(\mu-\nu)}\,
\frac{\Gamma(j-\nu+1)}{\Gamma(j-\mu+1)}\,
\int_{-\infty}^{\infty}d\alpha\,e^{-\mu\alpha}\,~~~~~~~~~~~~~~~
$$
$$~~~~~~~~~~~~
\times(\sqrt{z^2-1}+z\cosh\alpha+\sinh\alpha)^{-(j-\nu+1)/2}\,
$$
\begin{equation} ~~~~~~~~~~~~~~
\times(\sqrt{z^2-1}+z\cosh\alpha-\sinh\alpha)^{-(j+\nu+1)/2}\,.
\label{Qrep}
\end{equation}

The relations (\ref{PQa-}) and (\ref{PQa+}) allow us to write
the addition theorem directly in the form of a series:
$$
e^{-i\mu\theta_2}\,Q^j_{\mu\nu}(z_\theta)\,e^{-i\nu\theta_1}
=\vartheta(z_1-z_2)\sum_{\lambda-\nu=n}(-1)^{\lambda-\nu}\,
Q^j_{\mu\lambda}(z_1)\,P^j_{\lambda\nu}(z_2)\,e^{i\lambda\theta}~~~~~
~~~~
$$
\begin{equation}
~~~~~~~~~~~~~~~~~~+\vartheta(z_2-z_1)\sum_{\lambda-\mu=n}(-1)^{\mu-\lambda}\,
P^j_{\mu\lambda}(z_1)\,Q^j_{\lambda\nu}(z_2)\,e^{i\lambda\theta}\,.
\label{Qzt}
\end{equation}
In (\ref{Qzt}) $\,n$ runs through all integer values, and $\vartheta(x)$
is the usual discontinuous function: $\vartheta(x)=1\,$ for $x>0$ and
$\,\vartheta(x)=0\,$ for $x<0\,$. It is easy to write down one further
addition theorem:
$$
e^{-i\mu\theta_2}\,P^j_{\mu\nu}(z_\theta)\,e^{-i\nu\theta_1}
=\vartheta(z_1-z_2)\sum_{\lambda-\nu=n=-\infty}^{+\infty}(-1)^n\,
P^j_{\mu\lambda}(z_1)\,P^j_{\lambda\nu}(z_2)\,e^{i\lambda\theta}~~~~~
~~~~
$$
\begin{equation}
~~~~~~~~~~~~~~~~~~+\vartheta(z_2-z_1)\sum_{\lambda-\mu=n=-\infty}^{+\infty}
(-1)^n\,P^j_{\mu\lambda}(z_1)\,P^j_{\lambda\nu}(z_2)\,e^{i\lambda\theta}\,.
\label{Pzt}
\end{equation}

It is not hard to verify that (\ref{Qzt}) is equivalent to the analytic
continuation of (\ref{Qlam}) to imaginary values of $\alpha\,$.\footnote{
Even easier is to continue Eq. (\ref{Pzt}) into the region $\,-1< (z_1,
\,z_2)<+1\,$. Then, if all the indices $j,\,\mu,\,\nu$ are either integer
or half-integer with $\,-|j+1/2|<(\mu,\,\nu)<|j+1/2|\,$, the continued
relation appears to be equivalent to the known addition theorem for the
functions $d^j_{\mu\nu}\,$.} The addition theorem for $P^j_{\mu\nu}\,$
can also be written in integral form. It is, however, much more cumbersome
than (\ref{Qlam}) and we shall not write it out here.

In conclusion we note the following. As can be seen from the definitions
(\ref{defP}) and (\ref{defQ}), the theory of generalized Legendre functions
is equivalent to the theory of hypergeometric functions. Therefore for
each relation for generalized Legendre functions there are some
corresponding relations for hypergeometric functions. In particular,
there are some rather complicated nonlinear relations between hypergeometric
functions which correspond to the addition and multiplication theorems.
The author does not know whether these relations
exist in the mathematical literature. In any case they are not in the
well-known reference books~\cite{Bat,GR}.

An important part in the accomplishment of this work has been played by
discussions with A. A. Ansel'm, V. N. Gribov, G. S. Danilov, and  I. T. Dyatlov,
and some of the formulas relating to generalized Legendre functions were derived
in collaboration with them.


$$~~~~~$$

Translated by W. H. Furry

\end{document}